\documentclass[sigconf]{acmart}

\usepackage{booktabs}
\usepackage{tabularx}

\setcopyright{acmcopyright}
\copyrightyear{2021}
\acmYear{2021}
\acmDOI{10.1145/3411764.3445301}
\acmConference[CHI '21]{CHI Conference on Human Factors in Computing Systems}{May 08--13, 2021}{Yokohama, Japan}
\acmBooktitle{CHI Conference on Human Factors in Computing Systems (CHI '21), May 08--13, 2021, Yokohama, Japan}
\acmPrice{15.00}
\acmISBN{978-1-4503-8096-6/21/05}

\begin{document}

\title{``I Might be Using His\ldots But It is Also Mine!'': Ownership and Control in Accounts Designed for Sharing}

\author{Ji Eun Song}
\affiliation{%
  \institution{Seoul National University}
  \city{Seoul}
  \country{South Korea}
}
\email{jieun.song@snu.ac.kr}

\author{Jaeyoun You}
\affiliation{%
  \institution{Seoul National University}
  \city{Seoul}
  \country{South Korea}
}
\email{you.jae@snu.ac.kr}

\author{Joongseek Lee}
\affiliation{%
  \institution{Seoul National University}
  \city{Seoul}
  \country{South Korea}
}
\email{joonlee8@snu.ac.kr}

\renewcommand{\shortauthors}{Ji Song et al.}

\begin{abstract}
A user's ownership perception of virtual objects, such as cloud files, is generally uncertain. Is this valid for streaming platforms featuring accounts designed for sharing (DS)? We observe sharing practices within DS accounts of streaming platforms and identify their ownership characteristics and unexpected complications through two mixed-method studies. Casual and Cost-splitting are the two sharing practices identified. The owner is the sole payer for the account in the former, whereas profile holders split the cost in the latter. We distinguish two types of ownership in each practice---Primary and Dual. In Primary ownership, the account owner has the power to allow others to use the account; in Dual ownership, Primary ownership appears in conjunction with joint ownership, notably displaying asymmetric ownership perceptions among users. Conflicts arise when the sharing agreements collapse. Therefore, we propose design recommendations that bridge ownership differences based on sharing practices of DS accounts.
\end{abstract}



\maketitle


\section{Introduction}

The recent coronavirus (COVID-19) has caused everyone to be socially distanced. However, it has led to virtual bonding among people. The practice of sharing accounts has become a part of our daily lives. Specifically, the account of subscription-based streaming platforms, such as Netflix, Hulu, Spotify, Tidal, and Nintendo Family are frequently shared. According to digital platform statistics~\cite{inglesant2010,iqbal2020}, the total number of worldwide subscribers of Netflix, Amazon Prime, and Hulu by the end of 2019 was 167 million, 150 million, and 32 million, respectively. The number of streaming platform users has risen exponentially since the onset of the COVID-19 pandemic, according to Deloitte's digital trend report~\cite{westcott2020} published in June 2020.

Many streaming platforms provide the multiple-profile feature, allowing several users to simultaneously log on legitimately on to a single account. Depending on the service, platforms permit up to seven individual profiles that distinguish users within an account. Obada-Obieh et al.~\cite{obada2020} termed this type of accounts as ``designed for sharing (DS).'' DS accounts allow individuals to manage not only the platforms themselves but also the contents provided through their designated profiles across any network-connected device of their choice. The number of people who share DS accounts is considerable. In a survey~\cite{kafka2019} of 5,721 residents of the United Kingdom and the United States aged 16 to 64, 65\% were sharing Netflix accounts with others.

In the act of sharing, there is a sharer and a sharee -- the \textbf{account owner (AO)} and the \textbf{profile holder (PH)}, respectively. The AO, the sharer, can also be a PH, but as the AO creates the account, they decide who joins the DS account. The PH, the sharee, gains access credentials from the AO. Streaming platforms do not distinguish these user types, giving both the sharer and sharee comparable account control in order to minimize their control differences: anyone with the log-in information can access settings and the other profiles, except for payment information.

The question of ownership arises with multiple users being added to the DS accounts. Odom et al.'s~\cite{odom2012,odom2014} study about the usage behavior of virtual objects and possessions, suggested an uncertain recognition of ownership in cloud platforms due to the presence of platforms' properties; users can connect to clouds simultaneously, through various devices and from multiple locations, without intruding on the physical space of others. Furthermore, streaming services also attempt to limit users' ownership through legal, business, and technical means, as can be seen in End-User Licensing Agreements (EULAs), Terms of Service (ToSs), and Digital Right Managements (DRMs)~\cite{perzanowski2016}. Thus, the nature of ownership in streaming platforms may not be conspicuous in terms of the ability to use the platform despite the existence of multiple users.

Nevertheless, the perceived ownership of DS accounts is more noticeable than that of other virtual objects. Specifically, its complication lies both in the account credential of a particular user and the collective usage of virtual objects. Generally, the access credentials are intertwined with the owner's identity as the name and email address are submitted while generating an account. In this case, a sole user has ownership~\cite{gruning2016}, which gives a sense of control through self-identity~\cite{pierce2001}. However, in the DS accounts, the same level of access is granted to all PHs as their profiles are created with evenly divided boundaries. With PHs joining the account, the ownership perception may appear as ``mine'' or ``yours'' through relative identity~\cite{rader2009} as all members are allowed and perceived to have an equal share of ownership~\cite{gruning2016}. However, the ownership not recognized as in forms of ``ours'' can lead to a lack of accountability in PHs, creating a potential flaw in managing shared accounts, especially when tensions arise.

Therefore, this study examines the impacts of ownership perception has on its account control when multiple users share streaming platform DS accounts. The questions we explore are:

\begin{enumerate}
  \item How do users share DS accounts? With whom do they share the account, and what is the purpose of their sharing?
  \item How is the ownership perceived about accounts and among sharing users? How is it manifested?
  \item What problems does ownership perception cause in the extent of control of an account?
\end{enumerate}

This study used a mixed-method approach to gain insights into the prevalence of DS account sharing and to understand its perceived ownership and control. Specifically, this study conducted a descriptive online survey with 160 participants sharing accounts on streaming platforms in South Korea to discover the prevalence of DS accounts sharing, and a semi-structured interview with 31 users sharing 13 accounts in total to understand the sharing phenomenon.

This study's significance lies in its new discoveries about aspects of sharing-induced accounts and the in-depth analysis of problems caused by varying types of ownership. We find that sharing practices are either \textbf{Casual} or \textbf{Cost-splitting}. Casual sharing is relationship-driven without a clear purpose, whereas Cost-splitting sharing occurs regardless of the relationship among the shared users and solely to split costs. Casual and Cost-splitting sharing demonstrate two types of ownership: \textbf{Primary ownership} and \textbf{Dual ownership}, respectively. With Primary ownership, an individual account owner allows others to use the account, whereas Dual ownership is a combination of primary and joint ownership where all users use the account simultaneously with an equal share of the ownership. However, the coexistence of primary and joint ownerships in Cost-splitting sharing leads to an imbalance in control. At its worst, the asymmetry may result in the collapse of the ownership agreement when problems emerge. Based on the findings, this study discusses the need for an enhanced socially translucent system that ensures equal access to transparency among users. The application of this novel sharing design and the knowledge of differences in ownership perception that affects the account control can better equip streaming platforms to weigh in user experience in sharing, by promoting shared ownership while also maintaining individual ownership.

\section{Related Work}

We base our research on prior studies of account sharing in Human-Computer Interaction (HCI), ownership perception of digital objects, and controls within a multiuser environment.

\subsection{Account Sharing Practices in HCI}

Traditional consumer studies have analyzed subjects and their relationships with others with whom they share commodities, as well as motives and aspects of the sharing practice. While there is considerable debate in consumer studies regarding sharing, Belk~\cite{belk2007} defines the act of sharing as ``the act and process of distributing what is ours to others for their use and/or process of receiving or taking something from others for our use.'' Shared objects include physical goods, space, knowledge, and more~\cite{hellwig2015}, motivated primarily by altruistic~\cite{lampinen2014} or functional reasons~\cite{belk2014}.

HCI studies have focused on the device layers of the sharing phenomenon or the motives and aspects of sharing, as well as on problems in a socio-cultural context~\cite{busse2013,james2007,murphy2011,muller2012,karlson2009,obada2020}. In these studies, account sharing has been observed considering one single profile. A single profile account is usually shared in a dichotomous class of physical space either at work or in a non-work setting~\cite{adams1999,blythe2013,inglesant2010,kaye2011,koppel2015,lampinen2014,matthews2016,olckers2012,perzanowski2016,sasse2001,singh2007,steenson2008}. Sharing in households stems from convenience and is supported by trust, while sharing in a workspace, where it has a clear purpose such as achieving tasks, is based on utility in addition to trust~\cite{bartsch2013,frohlich2003,grassi2017,hassidim2017,koppel2015,matthews2016,song2019,weirich2001}. Account credentials are shared between participants in certain relationships; this sharing is often seen between couples, where an account is shared to support and maintain the relationship or promote intimacy and to keep track of what the other person is doing~\cite{bevan2018,jacobs2016,park2018}.

Although not common, even within a setting where multiuser profiles can be employed, usage may be limited to one device. Egelman et al.~\cite{egelman2008} investigated the mutual use of desktops in households and the multiple profiles setting application. Family members had trouble representing multiple people in a single profile and struggled to customize settings and contents, expressing difficulty with the time cost and mental effort of switching among profiles. As an alternative, a new form of account called a ``family account,'' a compromise between a single shared profile and individual profiles for each family member, was suggested.

Matthews et al.~\cite{matthews2016} noted that among the six types of sharing observed in a household -- borrowing, mutual use, set-up, helping, broadcasting, accidental -- mutual use typically occurs on an entertainment platform or desktop that has DS accounts. A distinct sharer and sharee may not necessarily exist; instead, two users have similar levels of ownership of the device or account. However, the dynamics of mutual use have been shown to be complex depending on various aspects of sharing, and it is not easy to see equal or level ownership on this type of account or profile. Thus, this study sought to observe ownership aspects in sharing an account with others, especially ownership that is recognized after the act of sharing.

\subsection{Ownership Perception of Digital Objects}

Although ownership is traditionally a legal concept~\cite{snare1972}, it has also been considered a psychological concept~\cite{beggan1992}. This study concentrates on the psychological side of ownership because no user legally owns the streaming service content. Wilpert~\cite{wilpert1991} explained that ownership answers the question of ``what do I feel is mine,'' since psychological ownership is based on a formed sense of possession toward an object. Built on consumer studies researchers Pierce~\cite{pierce2001} and Avey et al.~\cite{avey2009} and the suggestions of Olckers and Du Plessis~\cite{olckers2012} about components of ownership, Kuzminykh~\cite{kuzminykh2020} summarized psychological ownership from an HCI perspective in five dimensions:

\begin{itemize}
  \item \textbf{self-identity:} possession of the target becomes a ``representation'' of an owner,
  \item \textbf{self-efficacy:} the owner's judgment of his/her capability and competence to perform a task and to control the target,
  \item \textbf{accountability and responsibility:} the voluntary or enforced authority and obligation to take care of the target and related performances, consequences, and issues,
  \item \textbf{territoriality:} the owner's identification of the possession through external references leads the owner to defend the target if ownership or autonomy is endangered,
  \item \textbf{autonomy:} the owner's judgment of his/her capability to initiate decisions and actions regarding the target independently.
\end{itemize}

We categorize these ownership dimensions as autonomy-oriented and accountability-oriented. Self-identity, autonomy, and self-efficacy are \textbf{autonomy-oriented}, and accountability and responsibility, territoriality, and autonomy are \textbf{accountability-oriented}.

For devices with tangible characteristics, the physical location has represented psychological ownership. McMillan et al.~\cite{mcmillan2015} claimed that ``many of our concepts of digital media still draw directly on models of interaction developed when media had a physical instantiation.'' Hughes et al.~\cite{hughes1998} described the relationship between technology use and space ownership within the home: occupants used technologies such as the television to indicate that they controlled behavior in that part of the home. Frohlich and Kraut~\cite{frohlich2003} studied computers' position in the home by observing the relationship between computer location and sharing. They concluded that putting computers in private spaces gives special privileges to the owner of the space and discourages sharing while placing computers in a more public space encourages sharing.

Conversely, psychological ownership of virtual objects is less recognizable due to the inherent nature of virtual things -- placelessness, spacelessness, formlessness -- as Odom et al.~\cite{odom2014} stated. Virtual objects' psychological ownership diminishes as it moves from a file to a cloud server. Marshall et al.~\cite{marshall2012} examined to confirm ownership awareness between a service provider and cloud owner through an early cloud research study; the study established that people showed no interest in such a distinction or did not carefully consider the difference. Through a comparative cultural study of virtual possession of young adults, Odom et al.~\cite{odom2013} concluded that users' sense of ownership of streaming content was uncertain due to living in ``unfinished'' spaces and that they often experience a sense of fragmentation when trying to integrate their virtual possessions into their lives.

Especially with non-material objects, recognition of simple sole ownership~\cite{furby1980} and joint ownership~\cite{belk2012} wanes. Gruning et al.~\cite{gruning2016} studied user awareness of digital possessions such as pictures and music files stored on hardware devices and found that files in a shared family PC were typically considered jointly owned by several people rather than primarily owned by an individual who allowed others to use them. On the other hand, no real perception of ownership was observed with streamed content, and there was no expectation of long-term, persistent access to that content. In this study, we referenced the various dimensions of ownership compiled by Kuzminykh and analyzed how an AO and PH perceive joint ownership.

In sum, existing studies have shown that there are different types of shared ownership, but the possible conflicts arising from different perceptions of ownership held by sharers has not been explored. Our work addresses this gap by contributing a concrete understanding of the asymmetric perception of ownership among users and its potential to destroy an entire sharing agreement.

\subsection{Controls in a Multiuser Environment}

Based on the traditional HCI research approach of assigning one user per interface, sharing behavior encounters the problem of peer-to-peer control in multiuser interaction. Roger~\cite{rogers2011} argued that people naturally use technology in a shared way, even with devices designed for individual users. Many existing identity guidelines~\cite{grassi2017} cite security issues as the most significant reason for this tendency of shared use. By default, control studies of multiuser interactions have primarily addressed domestic technology and situations where multiple users are forced to interact. Considering roles and privileges in such a setting, Jang et al.~\cite{jang2017} classified users as primary, alternate primary, or guest. For desktop use, which is a good representative of an interactive home system, O'Brien et al.~\cite{obrien1997} concluded that users were not necessarily happy having expanded functionality in a single device; the researchers explored many of the social issues involved. Through research on control of access to digital contents within households, Mazurek et al.~\cite{mazurek2010} found that users created their own access-control mechanisms when sharing space on a single desktop. For example, they distinguished between reading and writing access in people's physical or remote presence and between locations such as at home or in public.

As noted earlier, HCI studies have focused on single-profile type accounts or on shared design limited to specific devices or spaces only, thereby overlooking control issues. As a result, technology control in multiuser interaction has been characterized as adopting an all-or-nothing approach. Karlson et al.~\cite{karlson2009} evaluated an all-or-nothing access model related to shared usage of personal phones. They found that the model does not support the range of user needs that they observed. Ur et al.~\cite{ur2013} also observed an all-or-nothing access model applied to domestic technology and found that such approach fails to provide seemingly essential affordances. In response, Cecchinato et al.~\cite{cecchinato2017} argued that this method of control creates a mismatch between users and owners in terms of the degrees of agency in users of home IoT devices, leading to a drop-in opportunity for behavioral affordance. Researching controls within interactions in multiuser smart homes, Geeng et al.~\cite{geeng2019} found that even within a multiuser platform when the primary user is absent, other users have difficulty controlling the account. The study, therefore, claimed that agencies should be differentiated for access to smart devices and emphasized the importance of considering power difference minimization implications. Zeng et al.~\cite{zeng2019} similarly claimed that multiuser, multi-device systems could face unique security, privacy, and usability changes, suggesting an application with concrete features such as location-based access controls, supervisory access controls, and activity notification as a solution.

Based on the inherent nature of virtual objects as defined by Odom et al.~\cite{odom2012}, an account can be considered as a digital object where awareness of management, users' accountability, and control are all weakened. In their study of user password practice, Adam and Sasse~\cite{adams1999} concluded that participants lacked security motivation and understanding of password policies and tended to circumvent password restrictions for the sake of convenience. Song et al.~\cite{song2019} pointed out the problems of account sharing in workplaces; tensions can arise from a lack of accountability, as can conflicts over simultaneous access and control. In fact, many users share account access with others without knowing that the account is being shared. According to Shay et al.'s~\cite{shay2014} study of account-sharing behavior, 30\% of 294 participants had an email or social networking account accessed by an unauthorized party. The decision to terminate an account is under one's control, but some subjects experienced cognitive and social-psychological burdens ending the sharing of their own accounts~\cite{obada2020}. Therefore, there are limits to applying various means of minimizing the account-authority gap on existing devices. Based on design suggestions, DS accounts are configured in a way that minimizes power differences. Despite these strands of research, little is known about perceived ownership of DS accounts or users' behavior when regarding power-difference minimization. Our work contributes a more concrete understanding of users' sharing practices and asymmetrical control issues, explaining how this knowledge could support more substantial ownership that could lead to better control of the account.

\section{Methods}

We conducted a two-phase study of user perception of ownership on streaming platforms with DS accounts in South Korea and the associated control issues. The study consisted of an online survey of 160 participants and 31 semi-structured, in-depth interviews. According to Blom et al.~\cite{blom2005}, the analysis of individual cultures can help identify universal behavior patterns. Further, following Hofstede's~\cite{hofstede2011} theory on Individualism versus Collectivism in cultural dimensions, we believe that this population set provides a starting point for the generalization on the sharing of technology and ownership perception in a collectivist cultures, despite its limitation in reflecting a specific milieu.

\subsection{Online Survey}

The purpose of the online survey was to gather a background understanding of sharing usage. In our three-minute survey of short-answer questions, we asked about DS account-driven streaming platforms in participants' current subscriptions and about how participants ended up in that platform. Participants selected the platforms they used from a list of streaming platforms we provided.

We recruited participants by posting an online advertisement offering a gift card compensation worth \$1. Users over 18 years of age were recruited based on three screening factors: 1) users (AO or PH) who had been subscribed to the platform for more than five months as of the date of the study; 2) users who were planning to continue their platform subscription; and 3) users who could verify their answers about the number of profiles associated with the account through a screen capture of the most frequently used account profiles. A usage of five months was determined to be key, as we were skeptical that shorter subscription periods could result in lower levels of account sharing and undeveloped sharing usage between users and the AO.

We calculated descriptive statistics for the average number of platforms. We used a general inductive approach for the free-form responses by collaboratively developing a coding scheme into two general themes. The study's primary author defined the initial codebook, and two independent coders categorized each response.

\subsection{Semi-Structured Interview}

In addition to the online survey, we conducted semi-structured interviews with 31 users (m=16, f=15) sharing 13 accounts in total. The participants consisted of 13 initial online survey participants and an additional 18 participants recruited through snowball sampling. Of the initial online survey participants who agreed to participate in the semi-structured interview, 13 were selected based on their sharing backgrounds and most often used streaming platforms. Each participant interview lasted one to one-and-a-half hours.

Questions in the semi-structured interview included those concerning: 1) background for sharing, including how the members of the account were gathered and how it was decided who would join the account; 2) current viewing behaviors, including viewing hours and location, as well as device type; and 3) maintenance of account settings and control strategies, including complications from sharing. An informed consent form was provided to all interview participants. As compensation, we paid \$17 with an extra \$3 per additional participant referral.

As we identified patterns in participants' levels of ownership after investing some time in our research, follow-up questions that went into further detail were included in the interview protocol~\cite{patton2001}. Rather than questioning participants directly about ownership, we indirectly asked how the user experienced ownership based on Kuzminykh et al.'s~\cite{kuzminykh2020} study of ownership dimensions. At the interview's end, we asked whether users considered themselves the owners of the account. Depending on an individual's role within shared accounts, interview questions were modified, added to, or omitted as needed. In line with Geeng et al.'s~\cite{geeng2019} finding that the sharer is more likely to participate actively than the sharee due to the nature of multiuser technical receptivity, this study encouraged the participation of elusive sharees specifically to obtain an in-depth understanding of sharing usage. We first considered AO recruitment of PHs and then asked the PHs about the recruitment process and if they knew the other PHs personally.

Interviews were conducted on Zoom due to pandemic restrictions. Table~\ref{tab:participants} summarizes the characteristics of our 31 participants, who ranged from 22 to 53 years of age. The participants' occupations varied and included housewife, undergraduate student, sales manager, factory manager, concert director, medical doctor, freelancer, public officer, furniture designer, TV scriptwriter, lawyer, and engineer.

Analyzing the interviews, we iteratively coded each interview using a bottom-up approach based on field notes and transcriptions. The first round of coding was conducted with an open coding process. The process was collective; authors analyzed transcribed interviews, initially categorizing significant responses into an incipient list of themes. The authors validated the theme schemes by working together and revisiting previously analyzed transcripts as they discussed the collected data with iteration, reaching a consensus about the codes matching the themes. As the process evolved, coding schemes were broadly formulated as sharing usage, ownership perception, and tensions. We used the percentage agreement metric described by Graham et al.~\cite{graham2012}. The result of the intercoder reliability test showed a strong agreement (Cronbach's $\alpha = 0.77$).

Throughout the study and this paper, we referred to participants using a naming scheme that identified their sharing usage and user type: C for Casual, and CP for Cost-splitting; O for AO, H for PH, and L for the lendee in a single profile. For example, C1O signifies that the participant is the owner of Casual sharing number one.

\subsection{Casual Sharing vs. Cost-splitting Sharing}

Casual sharing users exhibited an altruistic incentive in sharing their accounts. On the other hand, Cost-splitting sharing users were more focused on the practical issue of subscription costs. Thus, the payment methods of Casual and Cost-splitting can be characterized as non-split and cost-split, respectively. Non-split reflects the AO as the sole contributor to subscription fees, while cost-split usage expects equal contributions from all users.

\subsubsection{Casual Sharing}

Casual sharers demonstrated an altruistic motivation for account sharing, usually with family or friends. Four out of five owners in the Casual sharing account echoed participant C4O's opinion on sharing. He stated that ``the more money my family saves, the more I save as well.'' The AO had no obligation to share the account with PHs, yet still did so with the belief that they would benefit from sharing.

Casual sharing accounts were paid for by the AO primarily, and, in most cases, account credentials were shared by chance, often after discussing the content of a particular streaming platform. This process differed from when the sharer selectively recruited a sharee. C2H1 recounted, ``Our team has a new boss, and we were talking about Netflix one day. I told him I was planning on creating an account, to which he said he had a spot open and that he could share it for free.'' Some AOs assigned a person other than themselves to pay the subscription fee and offered to share the account with the payer afterward. For example, CP3O stated, ``This account belongs to me, but I'm paying with my mom's credit card and shared the account with her just in case she wanted to use it.'' Unless the relationship between an AO and the PH(s) went awry, a consistent group of PHs use the account until the AO ends the subscription. C1O noted, ``I've been sharing my account for 19 months now. It's awkward to take my offer back, nor does it really affect me.''

We also observed that AOs do not create a separate group chat with the PHs for account management. Instead, they prefer to either individually message the PHs or communicate face-to-face. ``There isn't much to manage,'' one user noted. ``I once messaged my sister the account password because she changed her phone and needed to log back in.'' (C5O)

\subsubsection{Cost-splitting Sharing}

For Cost-splitting sharers, the incentive to share was reducing the individual cost of the subscription. All AOs and PHs of Cost-splitting sharing, a total of 17 participants, fell into this category, citing the benefit of ``penny-saving'' in their motivations. CPH1 stated, ``It basically costs the same as a cup of coffee.'' Others mentioned the growing number of platforms that demand subscriptions as the source of the financial burden. For example, CP3O said, ``I am subscribed to four services because of my job: Netflix, Wavve, Apple Music, and TVing (video streaming platform). If I paid for all the subscriptions myself, they would cost me over \$50 per month.''

These users split the cost evenly among those who shared the account. Unlike Casual sharers, Cost-splitting sharers must actively recruit sharees based on a relatively rigorous selection process. Sharers primarily considered privacy and convenience, ultimately selecting either acquaintances or strangers as cost-splitting sharees. When recruiting among acquaintances, the AO selection process was divided into two main patterns: the sharee's subscription availability and sustainability.

First, the sharee's subscription availability indicated a person's ability to join the AO's account when the opportunity was offered. Availability was due to the lack of a current subscription or the willingness to cancel an existing one. CP6H1 said, ``I once asked people in my department group chat if they were interested in sharing a Netflix account. Those who replied are the ones I currently share the account with.'' Sustainability also played a significant role in determining sharees. The PH had to show the AO interest in the platform's content, as well as in continued utilization of the platform, thus sustaining their status as a PH. ``I wanted to share a Netflix account with my friends, but I knew that none of them would enjoy watching it in the first place. So, I sought interest on my online school bulletin.'' (CP1H1)

Interestingly, our data revealed that some users preferred sharing with strangers. This preference generally arose from difficulties finding acquaintances that met the conditions of availability and sustainability. Moreover, some noted that they wished to avoid potential discomfort that could result from changing the boundaries of the existing relationship. ``I don't like the idea that my co-workers or friends can log into my profile and see what I have been watching or that I have to make sure everyone pays each month. I don't want to make the people I know feel uncomfortable over the little things.'' (CPH2)

The fact that PHs were easily replaced in a shared account stood out. Generally, the criteria for replacement were determined by the PHs rather than an AO. PHs of a Cost-splitting sharing account were more likely to leave and/or be replaced than in a Casual sharing account. CP5O said, ``One of the sharees left our subscription account, so I had to find another person online.''

As a loosely defined administrator, AOs recruited users through various means, both online and offline. It was observed that an AO's relationships with the PHs were not weighed heavily in the recruitment process. If it occurred offline, member recruitment took either a direct form, where the AOs sought out potential members from within their existing networks, or an indirect form, where potential members were referred from outside an AO's personal networks. CP3O noted her direct recruitment experience: ``I just asked some people I knew if they were interested. One of them was in the same class as me, another one I knew through the church, and the last person I met at the gym.'' The two forms were not mutually exclusive, and the indirect form often took place in conjunction with the direct form. This ``snowballing'' method extended member recruitment to people in the networks of the PHs that had been recruited previously. CP4H1 had never met the AO, but he and a friend from his club soccer team had been sharing an account with the friend's acquaintance for more than a year and a half. At the time of our interview with CP4H1, he still had not met the AO in person.

\begin{table*}[!t]
\caption{Characteristics of semi-structured interview participants}
\label{tab:participants}
\footnotesize
\begin{tabularx}{\textwidth}{@{}>{\raggedright\arraybackslash}p{0.07\textwidth}>{\raggedright\arraybackslash}p{0.13\textwidth}>{\raggedright\arraybackslash}p{0.11\textwidth}>{\centering\arraybackslash}p{0.08\textwidth}>{\raggedright\arraybackslash}p{0.22\textwidth}>{\raggedright\arraybackslash}X>{\raggedright\arraybackslash}p{0.11\textwidth}@{}}
\toprule
Account & Owner (O)/ Holder (H)/ Lendee (L) & Platform & Length of Use (months) & Logged Device & Other users recognized by Participants & Total Number of Account Users \\
\midrule
C1 & O & Wavve & 20 & TV & H1, H2, H3, H4, OL1, OL2, H1L1, H2L1 & At least 9 \\
   & H1 & & 6 & TV & O*, H2*, H3*, H1L1 & \\
C2 & H & Netflix & 13 & Smartphone & O, H2*, H3*, H1L1 & At least 5 \\
C3 & O & Netflix & 15 & Desktop, Tablet PC & H1, H2, H3 & At least 4 \\
   & H1 & & 11 & Smartphone, Laptop & O, H1, H2, H3* & \\
C4 & O & Wavve & 6 & Smartphone, Laptop, TV & H1, H2, H3 & At least 4 \\
   & H1 & & 2 & Desktop, TV & O, H2, H3* & \\
C5 & O & Netflix & 19 & Smartphone, Laptop & H1, H2, H3 & 4 \\
   & H1 & & 16 & Smartphone, Laptop & O, H2, H3 & \\
C6 & H1 & Netflix & 7 & Smartphone 1, Smartphone 2, Desktop & O, H2*, H3*, H4*, OL1* & At least 5 \\
C7 & O & Watcha & 13 & Smartphone, TV & O, H1, H2, H3, H4, OL1, H1L1, H1L2, H1L3, H2L1, H2L2, H2L3, H3L1, H3L2, L3H3, H4L1 & At least 16 \\
CP1 & O & Netflix & 18 & Smartphone, Tablet PC & H1*, H2*, H3*, OL1, OL2 & 8 \\
    & H1 & & 6 & Smartphone & O, H2*, H3* & \\
    & H2 & & 12 & Smartphone, Laptop & O*, H1*, H3*, H2L1 & \\
    & H3 & & 12 & Smartphone, Laptop, TV & O*, H1*, H3*, H3L1 & \\
    & OL1 & & 12 & Smartphone & O, H1*, H2, H3* & \\
    & H2L1 & & 12 & Smartphone, Laptop & O*, H1*, H2, H3* & \\
CP2 & O & Wavve & 18 & Smartphone, TV & H1, H2*, H3*, OL1 & At least 6 \\
    & H1 & & 18 & Desktop, Tablet PC & O, H2*, H3*, H1L1 & \\
CP3 & O & Netflix & 24 & Smartphone, Laptop, Tablet PC & H1, H2, H3, OL1 & At least 5 \\
    & H1 & & 5 & Smartphone, TV & O, H2*, H3* & \\
    & H2 & & 10 & Smartphone, Laptop, TV & O, H1*, H3* & \\
    & OL1 & & 8 & Smartphone, Laptop & O, H1*, H2*, H3* & \\
CP4 & H1 & Wavve & 18 & Smartphone, TV & O*, H2*, H3, H1L1, H3L1 & At least 6 \\
    & H1L1 & & 15 & Smartphone, TV & O*, H1, H2*, H3* & \\
CP5 & O & Watcha &  & Smartphone, Laptop, TV & H1*, H2*, H3* & At least 4 \\
    & H1 & & 3 & Smartphone, Tablet PC, TV & O*, H2*, H3* & \\
CP6 & O & Netflix & 21 & Desktop & H1, H2, H3, H3L1 & 6 \\
    & H1 & & 21 & Smartphone, Laptop & O, H2, H3, H3L1 & \\
    & H2 & & 21 & Smartphone, Tablet PC & O, H1, H3, H2L1, H3L1 & \\
\bottomrule
\multicolumn{7}{l}{\textsuperscript{1}Users marked with * are strangers.}
\end{tabularx}
\end{table*}

\section{Findings}

\subsection{Survey Results}

Initially, 196 participants were recruited, with 160 participants selected for final analysis. The average age of our participants was 26.84 (SD = 6.51), and their ages ranged from 18 to 50 years (10s = 13, 20s = 96, 30s = 42, 40s = 6, 50s = 3). Sixty-nine participants were male, and ninety-one were female. Fifty-eight participants answered that they were the account owner, while fifty-six responded that they were friends or acquaintances of the AO. Thirty-one participants were family members of the AO and thirteen participants reported themselves as unknown to the AO.

\subsubsection{Prevalence of Streaming Sharing}

Our data indicated that most users of multiple-profile streaming platforms shared their accounts with others. In fact, 94.86\% (n=143) of the research participants responded that they shared their accounts. The average number of subscribed platforms was 1.9 (SD = 0.9). A sizable portion of the participants held shared accounts on video streaming platforms more than other types, such as music or entertainment. Participants rated Netflix the most used and shared platform (59\%), followed by Wavve (video streaming, 28\%), Watcha (video streaming, 20.5\%), FLO (music streaming, 7\%), and, lastly, Nintendo Family (entertainment, 5.1\%). We determined that the dominance of video streaming services can be attributed to the fact that they are more prevalent than other platforms, such as music streaming platforms, which are generally designed based on a single profile interface.

\subsubsection{Backgrounds for Sharing}

Based on our open coding analysis, we discovered two core sharing practices: Casual and Cost-splitting. As shown in Table~\ref{tab:practices}, 41.5\% of participants with shared accounts were classified under Casual sharing. These users typically shared living space or were presented with a sharing opportunity by chance. In sharing the account, they were primarily sharing with other PHs with whom they had personal relationships, usually family members or close acquaintances. Cost-splitting sharing comprised more than half (53.74\%) of participant responses. These users usually recruited other PHs through acquaintances or online communities to minimize the burden of subscription fees. Lastly, some responses could not be categorized by sharing practice. For example, 4.76\% of participants mentioned ``penny-saving'' in their interviews without indicating whether that was the motivation behind account sharing or a consequence.

\subsection{Casual Sharing vs. Cost-splitting Sharing}

Casual sharing users exhibited an altruistic incentive in sharing their accounts. On the other hand, Cost-splitting sharing users were more focused on the practical issue of subscription costs. Thus, the payment methods of Casual and Cost-splitting can be characterized as non-split and cost-split, respectively. Non-split reflects the AO as the sole contributor to subscription fees, while cost-split usage expects equal contributions from all users.

\subsubsection{Casual Sharing}

Casual sharers demonstrated an altruistic motivation for account sharing, usually with family or friends. Four out of five owners in the Casual sharing account echoed participant C4O's opinion on sharing. He stated that ``the more money my family saves, the more I save as well.'' The AO had no obligation to share the account with PHs, yet still did so with the belief that they would benefit from sharing.

Casual sharing accounts were paid for by the AO primarily, and, in most cases, account credentials were shared by chance, often after discussing the content of a particular streaming platform. This process differed from when the sharer selectively recruited a sharee. C2H1 recounted, ``Our team has a new boss, and we were talking about Netflix one day. I told him I was planning on creating an account, to which he said he had a spot open and that he could share it for free.'' Some AOs assigned a person other than themselves to pay the subscription fee and offered to share the account with the payer afterward. For example, CP3O stated, ``This account belongs to me, but I'm paying with my mom's credit card and shared the account with her just in case she wanted to use it.'' Unless the relationship between an AO and the PH(s) went awry, a consistent group of PHs use the account until the AO ends the subscription. C1O noted, ``I've been sharing my account for 19 months now. It's awkward to take my offer back, nor does it really affect me.''

We also observed that AOs do not create a separate group chat with the PHs for account management. Instead, they prefer to either individually message the PHs or communicate face-to-face. ``There isn't much to manage,'' one user noted. ``I once messaged my sister the account password because she changed her phone and needed to log back in.'' (C5O)

\subsubsection{Cost-splitting Sharing}

For Cost-splitting sharers, the incentive to share was reducing the individual cost of the subscription. All AOs and PHs of Cost-splitting sharing, a total of 17 participants, fell into this category, citing the benefit of ``penny-saving'' in their motivations. CPH1 stated, ``It basically costs the same as a cup of coffee.'' Others mentioned the growing number of platforms that demand subscriptions as the source of the financial burden. For example, CP3O said, ``I am subscribed to four services because of my job: Netflix, Wavve, Apple Music, and TVing (video streaming platform). If I paid for all the subscriptions myself, they would cost me over \$50 per month.''

These users split the cost evenly among those who shared the account. Unlike Casual sharers, Cost-splitting sharers must actively recruit sharees based on a relatively rigorous selection process. Sharers primarily considered privacy and convenience, ultimately selecting either acquaintances or strangers as cost-splitting sharees. When recruiting among acquaintances, the AO selection process was divided into two main patterns: the sharee's subscription availability and sustainability.

First, the sharee's subscription availability indicated a person's ability to join the AO's account when the opportunity was offered. Availability was due to the lack of a current subscription or the willingness to cancel an existing one. CP6H1 said, ``I once asked people in my department group chat if they were interested in sharing a Netflix account. Those who replied are the ones I currently share the account with.'' Sustainability also played a significant role in determining sharees. The PH had to show the AO interest in the platform's content, as well as in continued utilization of the platform, thus sustaining their status as a PH. ``I wanted to share a Netflix account with my friends, but I knew that none of them would enjoy watching it in the first place. So, I sought interest on my online school bulletin.'' (CP1H1)

Interestingly, our data revealed that some users preferred sharing with strangers. This preference generally arose from difficulties finding acquaintances that met the conditions of availability and sustainability. Moreover, some noted that they wished to avoid potential discomfort that could result from changing the boundaries of the existing relationship. ``I don't like the idea that my co-workers or friends can log into my profile and see what I have been watching or that I have to make sure everyone pays each month. I don't want to make the people I know feel uncomfortable over the little things.'' (CPH2)

The fact that PHs were easily replaced in a shared account stood out. Generally, the criteria for replacement were determined by the PHs rather than an AO. PHs of a Cost-splitting sharing account were more likely to leave and/or be replaced than in a Casual sharing account. CP5O said, ``One of the sharees left our subscription account, so I had to find another person online.''

As a loosely defined administrator, AOs recruited users through various means, both online and offline. It was observed that an AO's relationships with the PHs were not weighed heavily in the recruitment process. If it occurred offline, member recruitment took either a direct form, where the AOs sought out potential members from within their existing networks, or an indirect form, where potential members were referred from outside an AO's personal networks. CP3O noted her direct recruitment experience: ``I just asked some people I knew if they were interested. One of them was in the same class as me, another one I knew through the church, and the last person I met at the gym.'' The two forms were not mutually exclusive, and the indirect form often took place in conjunction with the direct form. This ``snowballing'' method extended member recruitment to people in the networks of the PHs that had been recruited previously. CP4H1 had never met the AO, but he and a friend from his club soccer team had been sharing an account with the friend's acquaintance for more than a year and a half. At the time of our interview with CP4H1, he still had not met the AO in person.

\begin{table*}[!t]
\caption{Practices categorized by sharing background}
\label{tab:practices}
\footnotesize
\begin{tabularx}{\textwidth}{@{}>{\raggedright\arraybackslash}p{0.14\textwidth}>{\raggedright\arraybackslash}X>{\centering\arraybackslash}p{0.11\textwidth}>{\centering\arraybackslash}p{0.16\textwidth}@{}}
\toprule
\shortstack[l]{Sharing\\Practice} & Example Answers & N (\%) & \shortstack[c]{Average Number of\\Profiles (sd)} \\
\midrule
Casual & - I share my Netflix account with my spouse\\
& - I pay and added my family on my account\\
& - Wanted to show it to my kids\\
& - My parents subscribed and set it on TV\\
& - My friend invited me to use his\\
& - Colleague at work said there were room available on her account & \shortstack[c]{57\\(41.5\%)} & 4.06(0.81) \\
Cost-splitting & - The subscription fee seemed a bit pricey for the numbers of videos I watch, so I wrote a post at the online community which I go to call people together\\
& - Saw a posting online of someone looking for people to share the account\\
& - My friend suggested to split the fee and use the account together & \shortstack[c]{79\\(53.74\%)} & 3.62(1.04) \\
Unable to classify & - My sister said she would pay for me\\
& - Started to share my account because it is a waste if I use it alone\\
& - At first, I used it alone, but my friend suggested that I should start sharing it with acquaintances. So we created a new account with the purpose of sharing.\\
& - My friend said she didn't watch much, so she suggested to share the account & \shortstack[c]{7\\(4.76\%)} & 3.43(1.51) \\
\midrule
TOTAL & & \shortstack[c]{143\\(100\%)} & 3.85(1.22) \\
\bottomrule
\end{tabularx}
\end{table*}

\subsection{Shared Ownership: Primary vs. Dual}

In both Casual and Cost-splitting sharing, members exhibited a sense of shared ownership of the account due to the act of sharing itself. However, members in Casual sharing accounts assumed a form of accountability-oriented ownership centered around the AO. In contrast, members in Cost-splitting sharing accounts manifested a more autonomy-oriented distribution of ownership, more equally among the PHs. The level of shared ownership was reflected in the self-efficacy and autonomy given to each user.

\subsubsection{Accountability-Oriented Primary Ownership}

In Casual sharing, accountability-oriented ownership centered around the AOs and their role as sole administrator. In other words, the AO held Primary ownership as the sharer, and the PHs as sharees adhered to the AO's expectations. The AO exercised ownership by managing and preventing conflicts among users and, in the process, the AO ensured the full benefits to his/her own account. C3O illustrated this sentiment, ``This account is mine, and I gave them access. I should not be inconvenienced.''

Although the AOs managed account access credentials in a way that guaranteed themselves the right to full access, they still offered some PHs limited self-efficacy. For example, some AOs would share the account with the PH not by giving them the log-in information but by setting it up on the PH's device themselves. Talking to an old friend about a series he watched on Watcha, C7O offered to share his account by setting it up on his friend's phone himself. He reasoned, ``If my friend were to log in on different devices, I might not be able to log in myself. I mean, he wouldn't do it, but just in case.'' Because the PH did not receive the account access credentials, the PH's access to the account was limited to one device. However, on that device, the PH had full access to the account as long as they remained logged in on that device.

\subsubsection{Autonomy-Oriented Dual Ownership}

Contrarily, Cost-splitting sharing accounts users exhibited autonomy-oriented ownership. The act of cost-splitting gave the PHs a sense of ownership in the form of guaranteed autonomy to access. ``I don't know what the rules are, but wouldn't it make sense that I have the right to access what I paid for? I might be using his\ldots but it is also mine!'' (CP4H1). This perception of ownership did not conform to the two types of shared ownership established in previous HCI studies -- primary and joint -- but rather manifested both types simultaneously. This ``dual'' ownership acknowledged that though the account was owned primarily by one individual (AO), it gave equal autonomy to all users (PH). This equality and autonomy to access provided PHs a sense of ownership, though shared. Under autonomy-oriented ownership, users acted with self-efficacy and autonomy, and with accountability and responsibility to a limited extent.

After examining the types and number of devices PHs chose to connect to their accounts, we observed a degree of self-efficacy in the PHs. As long as they shared the cost, the PHs set up their accounts on multiple personal devices, including smartphones, tablets, PCs, and laptops. They also saved the access credentials on these devices for easier access in the future. However, the set-up was not necessarily limited to personal devices only. Users also set up their accounts on communal devices, such as TVs. By doing so, they were able to co-watch with a non-PH third party. ``At home, I set up the account on the living room TV so that my mom and my sister could use it, which also led to my dad using it. Now my parents watch shows together, and I eventually made a separate profile for them.'' (CP1H3)

Users often exercised autonomy by sharing their access with third-party viewers, not through a separate profile but by giving access to their own profile. Sharing the access credentials was not always necessary; PHs often co-watched with a third party or saved the account credentials on communal devices. When the PHs did share the access credentials with a third-party user for convenience, they did not feel the need to let the AO or other PHs know. CP6H2 said, ``\ldots splitting the cost with my friends in fourths gives each of us a quarter of the autonomy to the account. Since I pay my share, it should not matter on which device I access my profile. What's the use in telling them? I share my account with my mom. As long as we don't use it at the same time, to others it would be the same as just me using it.'' Other participants, including CP1H1, CP1H2, CP1H3, CP3H1, CP4H1, and CP5H1 expressed the same sentiment as CP6H2.

\subsection{PHs' Shirk Accountability and Territoriality Facing Trouble}

According to our interview, Dual ownership among cost-splitting sharers created control conflicts between AOs and PHs. When conflict arose among users, PHs often relied on the AO for resolution. Consequently, the AOs described feeling burdened by the administrator role. Furthermore, when more severe or complicated issues arose, the AOs had to go beyond mere management, holding sole responsibility for the situation. In some cases, the implicit contract of shared ownership collapsed utterly. These complications demonstrated issues of accountability and territoriality.

Users of DS accounts generally encountered two types of control-related issues: ``light'' and less severe problems that only went as far as annoying the AO, and ``heavy,'' more severe issues that could even amount to property damage. The less severe problems required ensuring order and control within the platform. Since PHs lack both interest in and knowledge of controlling and managing user access, AOs were primarily responsible for maintaining user convenience. Some common examples of light issues included when a third party unacquainted with the AO acquired access credentials and crowded out the PHs and when there were recurring uncomfortable conversations about splitting costs. Few PHs seemed to care about the possibility that people other than themselves, would unknowingly share access credentials with third-party users. While individuals shared their own profile with others, they disregarded the possibility that others would as well, ``Don't people usually use their profile by themselves?'' (CP1O) The uncomfortable conversations were usually about payment collection, as people often missed payment deadlines and forgot the account numbers. PHs needed to be reminded regularly by the AO, placing the burden of management on the AO's shoulders. ``I have to remind them every six months. It's not that big of a deal, but it is sort of cumbersome to notify them that they have to pay every time.'' (CP2O)

Despite the amount of autonomy exercised under their partial ownership, PHs became completely dependent on their AOs when complications arose. In some cases, simultaneous access by unknown third-party free-riders precluded those who paid for the account from accessing it. These PHs, instead of checking account settings, went directly to the AO and requested help. CP1H3 recounted when he attempted to download videos from the streaming service that he was sharing with his family but could not. ``I wasn't exactly sure what was happening, but it seemed like there were too many devices accessing the account together. It makes no sense that I can't use what I pay for because of other people, though I guess that stuff like this happens when you share an account. Still, I contacted the AO because I didn't know what else I could do.'' Moreover, many PHs displayed passive attitudes toward payment and only paid when requested by the AO. One user noted, ``It's not a huge amount of money, so I tend to forget. Others remember to pay, and I also pay when I'm reminded. But it can get a bit awkward sometimes.'' (CP2H1) These users attributed their passivity to the fact that the AOs were the ones who created the account with their email addresses in the first place. Thus, the AO's nominal ownership of the account contributed to the PH's lack of perceived need to solve the problem themselves. ``Whatever the case, the AO is the one who created the account and pays the platform with their credit card, as well as who initially suggested splitting the costs.'' (CP6H3) When conflicts such as exceeding the access limit arose, although the source of the problem and necessary information to resolve it could be found easily, PHs would contact the AO first.

AOs often had no choice but to assume responsibility. AOs were solely in charge of policing account access when asked to do so by PHs, having to track and contact individual PHs. Unsurprisingly, AOs felt burdened by the role: ``If anything, it's annoying and burdening to have to manage other people's problems.'' (CP3O) The concentration of responsibility stemmed from the fact that AOs believed no one else was aware of how to access information about device-streaming activities. Their continued assumption of the administrative role resulted in them being the only ones knowledgeable about the account information. As the person who initially set up the account and recruited PHs, the AO was most familiar with checking device-streaming activities and even received email updates about them through the registered address. ``Whenever someone starts streaming, I get an email on their whereabouts. Since people use my email to log in, I'm the only one who sees what's going on.'' (CP2O)

Unsurprisingly, users preferred to take on the PH role to have less accountability in account management. The gap in general account information between PHs and AOs contributed to that preference. The asymmetry in the knowledge of account information weakened PH awareness of usage status. PHs had no interest in knowing about device activities because they never needed to know. In fact, PHs we interviewed did not even know that they had access to device-streaming activities and had never visited the settings page until the day of the interview. ``I had no idea that I could see other people's activities and whereabouts.'' We saw this response from both CP6H3 and CP3H1. Furthermore, these users were unaware that most streaming services provide a history of past activities for each profile on the settings page. ``I did not know until now that I had access to the entire access history!'' (CP6H2)

The more severe conflicts in account management exponentially widened the gap between an AO and the associated PHs in terms of responsibility. In an extreme case, an account on Wavve was hacked. ``I unexpectedly received an email that my account was accessed in Mexico and that the email address for log-in was changed'' (CP5O). CP4H1 was defrauded on Netflix: ``A week after I joined a shared account which I found online, the account disappeared.'' In these cases, all pre-established forms of ownership collapsed. Dual ownership became Primary ownership, as PHs pushed responsibilities to the AO. However, PHs were not unique in doing so; AOs also attempted to avoid taking control. For example, after being hacked, CP5O initially described a plan to inform PHs about the situation to remain in charge, but later decided to cancel the subscription entirely. In short, previous expectations of accountability and territoriality no longer applied. Both Primary ownership and Dual ownership were dissolved. Unsurprisingly, our interviews revealed that people who had experienced these types of conflicts decided to cancel the existing membership (CP5O) or to convert to a private account (CP4H1).

\section{Discussion}

Our findings present three significant contributions. First, we have offered a categorization of DS account sharing usages based on users' motivations, noting that DS account sharing may be financially motivated. Second, we discovered that shared ownership exhibits new ownership characteristics in the presence of multiple users. As the cost is split among users, a combination of Primary and Dual ownership of the account results; the AO holds dominance as a primary owner, at the same time, each user has an identical autonomy within the account. Finally, we have provided insight into potential ownership conflicts in autonomy-oriented ownerships. Our research indicates that autonomy-oriented ownership dimensions become more evident as the difference in the information accessible to each user decreases.

There are two sharing practices when it comes to sharing DS accounts: Casual and Cost-splitting. In the former, the account owner is the sole payer for the account, and in the latter, the profile holders split the cost. Respectively the ownership is displayed differently in each practice. We observed Primary ownership in Casual sharing where the account owner is the only decision-maker to allow others to use the account, whereas, in Cost-splitting, such Primary ownership appears in conjunction with joint ownership. The gap of ownership perception provides the potential to provide light to heavy problems, destroying the sharing agreement at worst.

Unlike typical devices that are tangible, perceptions of ownership on DS accounts appear to be more complicated. While tangible devices demonstrate a clear distinction between primary and joint ownership, this distinction becomes blurrier in shared ownerships with virtual objects. Often, users have different conceptions of their ownership, which also impacts the control and power distribution among users.

Our study began by investigating the phenomenon of asymmetrical control over shared accounts. In addition to analyzing the underlying reasons for asymmetrical control, we expanded our study to suggest an alternate design that promotes shared ownership while maintaining individual ownership. We suggested a platform design that 1) has an enhanced intermediary feature; 2) nurtures a sense of communal responsibility; and 3) supports a socially translucent system beginning on the first landing page users encounter on the streaming platforms. Although our research outcomes cannot fully prove our design recommendations' effectiveness, they nonetheless provide useful and applicable insights for streaming service users and providers.

\subsection{Enhancing Intermediary Features of the Online Platform}

Our study discovered a demand for a channel that connects users, yet the connection process is not without obstacles. Users who search for potential account sharers to save money must often conduct a laborious search accompanied by time and energy costs. Existing online platforms do not provide services that can sufficiently meet the need for connection without the time and energy difficulties.

The essential function of an online platform is to be an intermediary. Currently, platforms focus on connecting users with content, generating profit through subscriptions~\cite{srnicek2016}. However, our research suggests that online platforms should consider acting as agents that enhance peer-to-peer connections and, thus, increase user convenience, which could also lead to increases in profit. First of all, platforms can connect users for cost-splitting purposes. By providing the financial incentive of lower-cost, platforms can offer lowered barriers to entry and motivation for subscription continuation. We recommend that platforms consider serving as peer-to-peer mediums.

Streaming platforms can also provide user-matching services or run a separate platform focused on matching to reflect the diverse and evolving perceptions of ownership. Platforms could request access to a user's contacts -- phone numbers or social media accounts -- and display a list of people who are not yet subscribed but could join the user's subscription. Another useful contribution could be depersonalizing conversations about cost-splitting. For users logged in to the platform's app, the AO can set up app notifications that remind PHs about recurring payment dates.

We envision peer-to-peer services that would complement the platform's role as a content provider and enhance user convenience. By drawing in more people with the potential to become ``available'' and ``sustainable'' users, the platform could increase subscriptions. Users would appreciate the ability to share accounts more efficiently as well as the financial benefit sharing brings. Account sharing usually increases account sustainability due to lower costs and added social obstacles to leaving the account. While subscription sustainability was not a research topic of this study, we plan to investigate it in future research.

\subsection{Nurturing a Sense of Communal Accountability}

Our study has shown that PHs consume streaming services proactively but avoid situations in which accountability is demanded. They demonstrate proactiveness in searching for potential account sharers, setting up the account on their devices, and sharing their accounts with third-party users. Yet, once conflicts occur, they leave it to the AOs to resolve them.

Instead of conferring different levels of control and, thus, ownership to individuals, platforms should be designed in a way that encourages all users to be responsible for account management. According to Gruning~\cite{gruning2016}, ``different types of ownership could imply different sets of actions.'' Platform designs must reflect the diverse needs and desires of varying forms of ownership. However, the definition of ownership includes more than mere freedom and autonomy of use, as Gruning suggests; rather, it extends to assuming responsibility for management. Recent studies have demonstrated that gaps in power distribution among users should be minimized~\cite{geeng2019}. Yet, platforms still allow users to enjoy the freedom of access without responsibilities; thus, the PHs continue to be free-riders, and the AOs continue to be responsible for consequences.

Our findings indicated that reducing differences between levels of control power among users led to both the increase and diversification of sharing, sharing with friends and even strangers. Though most Terms of Services on streaming platforms stipulate that users are allowed to share their account with family members only, Cost-splitting users already share with a variety of people, including friends and even strangers. The only communal aspect in the current spectrum of shared ownership is that the users are under one account, but they are a group of relatively independent users who do not share very much other than this account access.

Providing a cost-splitting feature is one effective way to develop communal accountability within an account. In order to minimize the need for an AO to demand payments from the PHs, the platform could allow a direct charge of the split costs from each user. Therefore, the entire group would face the consequences of one user not paying on time. We acknowledge that this measure may simply shift the tension between the AO and PHs to among PHs. A possible solution would be to permit customization of payment deadlines for each user. Regardless of the exact set-up, this shift would initiate an effective transition into communal accountability. As the nature of the relationships between sharers and sharees inevitably differ by account, we suggest providing an option for account users to select their preferred method of cost distribution. For example, requiring families to pay separately would not be efficient or, likely, desirable. If a platform adopts a cost-distribution design, we expect it to lessen the management burden that the AO carries and increase the sense of control held by PHs. The effectiveness of a cost-distribution design can also be evaluated in future studies.

\subsection{Support for an Enhanced Socially Translucent System}

We suggest that streaming services strengthen inter-user transparency to increase awareness of one another in the shared account. Song~\cite{song2019} employed the socially translucent system of Erickson and Kellogg~\cite{erickson2000} to emphasize this need for awareness even for account sharing in the form of single profiles. Yet, in a current multiuser-account setting, account sharing in the form of single profiles provides limited to low accessibility in terms of others' viewing activity.

Social translucence is necessary because no user has the same mental model. If information is socially translucent to one particular person only, in this case, the AO, other users remain oblivious to the management needs of the account. This translucence challenge resembles concerns in past studies where users do not have a complete mental model with which to approach ownership if they are not made aware of the information collected and available.

Streaming platforms have made efforts to increase transparency among users by developing functions that inform about others' activities. For example, every user has access to activity history via account settings, along with time and location of access; the AO receives activity updates through the email address connected to the streaming account. Yet, we found that access to transparency is still limited to particular users, namely the AO. In fact, only AOs receive email updates and tend to be aware of the translucent system. Other PHs not only lack access to email updates but also have never visited the settings page. Similar to CP3H1's response, many PHs responded that they ``did not know [they] had access to such information because the account was not set up with their own email address.''

Thus, we propose that users be trained to access necessary account information outside the context of conflicts. We suggest displaying other users' activity on the common profile page that users see first when they utilize the platform. It would be important to include information such as the number of devices currently streaming on all logged-in devices on the page that acts as the first point of access. The objective is to encourage all users to become familiar with account usage status and learn about activity patterns. This strategy invites users to learn how to approach issues that arise and improves problem-solving efficiency. However, since this display arrangement may have implications for user privacy, we call for further examination of the extent of the information that should be shared. Additional research can be conducted on the topics of appropriate depth, and extent of user activity information shared within the account.

\section{Limitation and Future Work}

From a methodological point of view, our study is subject to some limitations in its research scope. One of the main limitations is with the subjects of the study, due to their technical backgrounds. This study's technical background centered on video streaming services more so than on game or music streaming services. The explanation for this choice is that there are not as many game or music streaming services available; the proportion of users is also significantly lower than video streaming users. Users of video streaming services equipped with multi-profile platforms were more readily available.

It is important to note that we did not investigate extensively users who were parasitically sharing a PH's profile instead of having their own profiles. We tried to recruit passive users, currently the chronic problem of traditional multiuser platforms. Through snowballing, we recruited profile holders who are conventionally considered as secondary users. However, as technology is shared, secondary passive users appear. We even found new passive users who were sharing one profile with an AO or PH, but it was difficult to recruit such users through snowballing, as in snowballing, the participants are not always able or willing to recruit others.

Another limitation is that the study was conducted with adult participants and did not fully address adolescents or children's usage behaviors. It is important to note that adolescents and children may also access credentials and may even share credentials with others.

This study examined user ownership and control when cost-sharing occurred with virtual objects. By considering the limitations mentioned above, we hope to explore in future research how sharing expansion in other cloud services appears. Future comparison of various sharing users' usage sustainability will allow us to examine PHs' ownership changes. To verify the design recommendation that this study suggests, another possibility is to observe the behavior changes when PHs are given the same information as the AO. Various subscription services are being introduced as the subscription economy expands. The differences in ownership concepts between virtual and material objects in these subscription services can also serve as new case studies.

\section{Conclusion}

Given the prevalence of sharing-inducing accounts in various streaming platforms, it is essential to look at how users accept and perceive ownership of, and at related complications of, these virtual objects. We conducted a mixed-method study, both quantitative and qualitative, to examine ownership-centered interactions and unexpected complications among sharing users. We found that platforms that are designed for sharing do induce users to share accounts. Initially, we observed two general sharing classifications: Casual and Cost-splitting sharing. We also discovered that the boundary of each user's ownership is relatively ambiguous in sharing-inducing accounts. This research contributes to a known gap in understanding ownership perception and, consequently, what is known about levels of control. Notably, we discovered that Dual ownership appeared in Cost-splitting sharing. This Dual ownership tends to display an asymmetric perception of ownership across different users that, at its worst, has the potential to destroy an entire sharing agreement. Therefore, we propose design recommendations based on everyday practices of account sharing and perceived ownership that bridge the ownership gap between users by suggesting an alternate design that promotes shared ownership while maintaining individual ownership.

\begin{acks}
We truly appreciate all the participants as well as reviewers of our study for their time.
\end{acks}

\bibliographystyle{ACM-Reference-Format}
\bibliography{references}

\end{document}